# Generalized Kappa Distribution Function for Mixed Fermiom-Boson Quantum Plasmas

Aakanksha Singh[1], Abhisek Kumar Singh[2] and Punit Kumar[1]
[1] Department of Physics, University of Lucknow, Lucknow, India
[2] Department of Physics, G L Bajaj Group of Institutions, Mathura, India

ABSTRACT

A Kappa distribution function applicable to systems comprising of mixed fermions and bosons has been developed through the thermodynamic Gibbs potential utilizing the quantum versions of the Olbert kappa distributions. The generalized expressions of the partition function and the entropy have been evaluated for such mixed quantum systems. The analysis shows that boson-rich systems consistently exhibit higher entropy, than fermion-rich systems. The distribution functions show heavy tailed characteristics at low Kappa values, indicating the presence of superthermal particles. It is observed that relativistic effects lead to significant increased entropy.

# 1. Introduction

The classical kappa distribution, first introduced by Stan Olbert in 1966 [1] and now referred to as 'Olbert's Kappa Distribution', has been pivotal in plasma physics and space science. The first one to apply this was Binsack [2], where he employed to study electron fluxes in Earth's plasma sheet [3]. Since its inception, the kappa distribution has been widely observed in particle distributions across high-temperature plasmas in space [4–10], with confirmed applications in diverse contexts such as cosmic ray spectra [11,12], the solar wind [13–17], shock regions [18–20], and the heliosphere [21]. Beyond classical applications, the kappa distribution has been extended to the relativistic domain and utilized in exploring statistical probabilities in correlated systems [22, 23]. Notably, its evolution under non-stationary conditions, particularly in particle acceleration scenarios, has been extensively studied. These investigations have contributed to the development of non-equilibrium statistical mechanics, where kappa distributions emerge as stationary states far from thermal equilibrium [24–28].

In nonlinear plasma theories, several classical distributions analogous to the kappa distribution have been proposed [29–31]. Within the framework of Tsallis' thermostatistics [32], the kappa distribution is considered a relative of Tsallis' non-extensive distributions due to their shared incorporation of entropy formulations that deviate from extensivity, commonly referred to as 'non-extensive' in Tsallis' theory. Recently, the construction of a super-extensive entropy [33], representing an analytical form of classical kappa entropy, has further expanded the theoretical understanding of kappa distributions, distinguishing them from traditional thermodynamic entropy. This advancement suggests that the fields of application for kappa distributions span distinct domains in physics and statistical mechanics.

Despite these developments, the application of kappa distributions to quantum systems remains a significant challenge. Existing kappa distributions, whether relativistic or non-relativistic, have primarily been limited to classical plasmas. Attempts to develop quantum analogs of classical kappa distributions have so far been inadequate, producing only simplified versions derived from the partition function [26, 34] that fail to capture the genuine quantum characteristics of these systems. Although Fermi-Dirac and Bose-Einstein distributions have been extended using Tsallis' non-extensive statistics [35–37], they diverge conceptually from Olbert's kappa quantum distributions. Recently, a more accurate quantum version of the kappa distribution has been formulated, encompassing Olbert distributions [38], with promising applications in quantum statistical mechanics. These quantum kappa distributions extend Olbertian statistical mechanics into the quantum domain and serve as a generalized analytical framework bridging Boltzmann-Gibbs (BG) statistical mechanics with Lorentzian systems. Under suitable conditions, this framework can describe both classical and quantum systems. However, applying these quantum kappa distributions to quantum plasmas necessitates a paradigm shift from the conventional fluid-based Madelung-Bohm theory to a statistical approach. Notably, the construction of a mixed Olbert-$\kappa$ distribution for systems comprising fermions and bosons remains an open research frontier.

In the universe, particles can be categorized into two classes, Fermions and Bosons, which encompass all the systems we encounter. This observation motivates the construction of a mixed Kappa distribution function. In this study, we develop the mixed Kappa distribution, a well-defined probability distribution that describes the two classes of particles (fermions and bosons) in their mixed state. The developed theory can be applicable to both quantum and classical plasmas i.e., by establishing the Kappa distribution function for mixed Fermion-Boson states, we can gain a deeper understanding of various types of plasmas. To construct the mixed Kappa distribution function for Fermion-Boson systems, we utilize the thermodynamic Gibbs potential of mixed states. This distribution function, alongwith the Gibbs potential and partition function, serves as the foundation for building a model that enables the study of relevant thermodynamic and statistical quantities. This extension of the Olbert-Kappa distribution may allows us to explore systems that exhibit mixed states.

In Sec. 2, the generalized expressions for mixed κ- distribution function, entropy and partition function have been derived. In Sec. 3, we present summary alongwith discussion.

## 2. Theoretical formulation

We start with the Gibbs potential, which is the simplest approach for deriving the distribution function. The Gibbs potential, being an energy quantity, remains additive (extensive) regardless of other properties of the gas. By fractionally combining the Olbert-Fermi and Olbert-Bose potentials, we obtain the Gibbs potential for the mixed case [38],

$$\Omega_\alpha^{OBF} = q\Omega_\alpha^{OF} + (1-q)\Omega_\alpha^{OB}, \tag{1}$$

where, $q<1$ is the non-extensive entropy's $q$-index and is related to Kappa-index (κ) by the relation; (q = 1+ 1/κ) [23]. The expressions for the Olbert- Fremi ($\Omega^{OF}$) and the complementary Olbert-bose ($\Omega^{OB}$) Gibbs potentials are given as [38],

$$\Omega^{OF} = \frac{\kappa}{\beta}\left[1 + \{1 + |\mu_f - \varepsilon_\alpha|\beta/\kappa\}^{-(\kappa+s)}\right]^{-1/(\kappa+s)} - \frac{\kappa}{\beta}, \ \mu_f > 0, \tag{2}$$

and

$$\Omega^{OB} = \frac{\kappa}{\beta}\left[1 - \{1 - (\mu_b - \varepsilon_\alpha)\beta/\kappa\}^{-(\kappa+s)}\right]^{1/(\kappa+s)} - \frac{\kappa}{\beta}, \ \mu_b \leq 0, \tag{3}$$

where, $\beta = T^{-1}$ (T is consistent physical temperature) and ($\varepsilon_\alpha$) is the energy of particular state α. $\mu_f$ and $\mu_b$ are the chemical potentials in case of fermions case and bosons respectively. κ is the Kappa index, s > 0 is a fixed constant number that is introduced to account for thermodynamic considerations. The specific values of this constant number are s = 4 and s = 5/2 for ideal relativistic gases and non-relativistic gases respectively, in accordance with the thermodynamic constraints [12, 35]. Using the thermodynamic potential, we can simply find the average occupation number which provides the expression for distribution function ($\langle n_\alpha \rangle = \frac{\partial \Omega_\alpha}{\partial \mu} \equiv P(\varepsilon_\alpha)$) [38]. By differentiating Eq. (1) with respect to chemical potential μ we obtain the required distribution,

$$\frac{\partial \Omega_\alpha^{OBF}}{\partial \mu} = q \frac{\partial \Omega^{OF}}{\partial \mu} + (1-q) \frac{\partial \Omega_\alpha^{OB}}{\partial \mu}, \tag{4}$$

and

$$\langle n_\alpha \rangle_{OBF} = q \langle n_\alpha \rangle_{OF} + (1-q) \langle n_\alpha \rangle_{OB}, \tag{5}$$

where, $\langle n_\alpha \rangle_{OF}$ and $\langle n_\alpha \rangle_{OB}$ are the Olbert κ-Bose and Olbert κ-Fermi distribution functions respectively [38] and the respective values are,

$$\langle n_\alpha \rangle_{OF} = \left\{ \left[ 1 + |\mu_f - \varepsilon_\alpha| \beta/\kappa \right]^{\kappa+s} + 1 \right\}^{-[1+1/(\kappa+s)]}, \tag{6}$$

and

$$\langle n_\alpha \rangle_{OB} = \left[ 1 - |\mu_b - \varepsilon_\alpha| \beta/\kappa \right]^{-2} \left\{ \left[ 1 - (\mu_b - \varepsilon_\alpha) \beta/\kappa \right]^{(\kappa+s)} - 1 \right\}^{-1/1/(\kappa+s)}. \tag{7}$$

Hence, the final expression for mixed κ- distribution function is

$$\langle n_\alpha \rangle_{OB} = q \left\{ \left[ 1 + |\mu_f - \varepsilon_\alpha| \beta/\kappa \right]^{(\kappa+s)} + 1 \right\}^{-[1+1/(\kappa+s)]}$$
$$+ (1-q) \left[ 1 - |\mu_b - \varepsilon_\alpha| \beta/\kappa \right]^{-2} \left\{ \left[ 1 - (\mu_b - \varepsilon_\alpha) \beta/\kappa \right]^{(\kappa+s)} - 1 \right\}^{-1/1/(\kappa+s)}. \tag{8}$$

Now, Entropy for the system having mixed state can be obtained by the negative derivate [$S = -\partial_T \Omega$] of the (mixed) Gibbs potential with respect to temperature [38] as,

$$S_\alpha^{OBF} \equiv \beta^2 \frac{\partial \Omega_\alpha^{OBF}}{\partial \beta} = \beta \left( \frac{\partial \beta \Omega_\alpha^{OBF}}{\partial \beta} - \Omega_\alpha^{OBF} \right). \tag{9}$$

Using Eq. (1), the above expression reduces to,

$$S_\alpha^{OBF} = q S_\alpha^{OF} + (1-q) S_\alpha^{OB}, \tag{10}$$

where, ($S_\alpha^{OF}$) and ($S_\alpha^{OB}$) are entropies for Fermi case and Bose case, respectively

$$\frac{S_\alpha^{OF}}{\kappa} = 1 - \sum_\pm \left\{ (1 + \lfloor \mu_f - \varepsilon_\alpha \rfloor \beta/\kappa)^{\pm(\kappa+s)} + 1 \right\}^{-1-1/(\kappa+s)}, \tag{11}$$

and

$$\frac{S_\alpha^{OB}}{\kappa} = 1 - \sum_\pm [1 - |\mu_b - \varepsilon_\alpha|\beta/\kappa]^{-2}\{[1 - (\mu_b - \varepsilon_\alpha)\beta/\kappa]^{\pm(\kappa+s)} - 1\}^{-1+1/(\kappa+s)}. \tag{12}$$

Thus, the expression for the (dimensionless) entropy of the mixed state α is,

$$S_\alpha^{OBF} = q\kappa\left[1 - \sum_\pm \left\{\left(1 + |\mu_f - \varepsilon_\alpha|\beta/\kappa\right)^{\pm(\kappa+s)} + 1\right\}^{-1-1/(\kappa+s)}\right] + (1-q)\kappa\left[1 - \sum_\pm [1 - |\mu_b - \varepsilon_\alpha|\beta/\kappa]^{-2}\{[1 - (\mu_b - \varepsilon_\alpha)\beta/\kappa]^{\pm(\kappa+s)} - 1\}^{-1+1/(\kappa+s)}\right], \tag{13}$$

and the corresponding partition function is

$$Z_{OBF} \equiv \sum_\alpha \langle n_\alpha \rangle_{OBF} = q\, Z_{OF} + (1-q)Z_{OB}, \tag{14}$$

where, $Z_{OF} = \sum_\alpha \langle n_\alpha \rangle_{OF}$ and $Z_{OB} = \sum_\alpha \langle n_\alpha \rangle_{OB}$ are the Olbert κ- Bose and Olbrt κ- Fermi partition functions respectively.

In the numerical analysis to follow, the parameters are chosen for dense astrophysical plasmas like white dwarfs and neutron stars, with typical values of $T \approx 10^5 - 10^8 K$, [39, 40] and the value of kappa parameter for such type of dense astrophysical objects ranges between 0.1 to 0.5 [41]. The specific values of constant '$s$' are taken to be 5/2 and 4 for ideal non-relativistic and relativistic gases [12, 35].

Fig. 1 shows the variation of entropy of a mixed fermion-boson system ($S_\alpha^{OBF}$) with kappa index (κ), for the particle densities ratio of fermion to boson $(\Delta = n_f/n_b)$. Lower κ values correspond to systems farther from equilibrium, resulting in higher entropy due to greater disorder and more pronounced non-equilibrium effects. As κ increases, the system approaches equilibrium conditions, and entropy decreases accordingly. Additionally, this curve is plotted for particle density ratio, demonstrating that this ratio of fermions to bosons influences the entropy profile. We found that for a fixed κ value, entropy is higher for lower values of $\Delta$, meaning boson dominated systems exhibit greater entropy than fermion dominated systems. This behaviour reflects the underlying quantum statistical nature of the boson that they can occupy the same quantum state following Bose-Einstein statistics. Especially at higher temperatures, where Bose-Einstein condensation hasn't yet dominated, bosons tend to have higher entropy than fermions, as they have more accessible microstates due to which they can bunch up in the same states, allowing many more configurations, while fermions constrained by the Pauli-exclusion principle contribute to reduction in entropy.

Fig. 2 shows the variation of the particle distribution function for mixed fermion-boson sysytem ($\langle n_\alpha \rangle_{OBF}$) against energy ($\varepsilon_\alpha$), for varying values of the kappa index (κ). At lower κ values, the distribution shows a heavy tailed behaviour, indicating a significant presence of high-energy particles (superthermal). This tail diminishes as κ increases, and the distribution transitions toward the ordinary form (Fermi and Bose distribution), which features exponential decay at high energies mirroring the behaviour of the Maxwell-Boltzmann distribution. The trend demonstrates how the mixed κ-distribution flexibly models the system with both low-energy particles (thermal) and high-energy particles (superthermal). This is particularly relevant in describing physical systems like dense astrophysical plasmas and condensed matter physics, where such mixed-energy distributions can be observed.

Fig. 3 shows the behaviour of mixed partition function ($Z_{OBF}$) as a function of temperature ($T$) for relativistic (solid line) and non-relativistic (dashed line) cases. As temperature increases, the relativistic partition function increases more rapidly compared to the non-relativistic one. The relativistic case shows non-linear, accelerated growth with temperature, reflecting the thermodynamic relationship where higher temperatures lead to a greater number of accessible energy states due to relativistic corrections affecting the density and distribution of energy states. In contrast, for non-relativistic cases, the partition function increases steadily i.e., a much slower, almost linear growth with temperature. This shows that relativistic effects significantly enhance the thermodynamic behaviour of a mixed quantum plasma system at high temperatures, leading to a higher partition function and consequently, higher entropy and energy capacity compared to the non-relativistic regime.

## 3. Summary and Discussion

In this work, we introduced a mixed Kappa distribution function to describe systems comprising both fermions and bosons, extending the classical Kappa distribution into the quantum domain. By utilizing the thermodynamic Gibbs potential for mixed fermion-boson states, we derived generalized expressions for the entropy and partition function of such systems. The effect of kappa parameter on entropy ($S_\alpha^{OBF}$) as well as on the distribution function ($\langle n_\alpha \rangle_{OBF}$), for the mixed system has been investigated while the effect of temperature on the mixed partion-function ($Z_{OBF}$) for two different cases, relativistic and non-relativistic has been also investigated. The generalized mixed Kappa distribution formulated in this study provides a robust framework for describing the non-equilibrium thermodynamics of quantum plasmas consisting of both fermionic and bosonic particles. The findings highlight the significant interplay between quantum statistical behaviour and non-equilibrium dynamics, with the Kappa index serving as a key parameter in modulating entropy and particle distribution characteristics. In mixed systems, bosons, owing to their ability to occupy the same quantum state, contribute more prominently to the overall entropy, particularly under non-equilibrium conditions. The presence of superthermal particle populations at lower κ values, and their progressive suppression with increasing κ, reflects realistic particle behaviour in dense astrophysical plasmas. Additionally, the sharp contrast between relativistic and non-relativistic thermodynamic response, represents the necessity of incorporating relativistic effects when analyzing high temperature quantum plasmas.

This theoretical framework not only encompasses the quantum versions of the Olbert-Fermi and Olbert-Bose Kappa distributions, but also provides a unified approach for studying the statistical and thermodynamic properties of mixed-state plasmas. The mixed kappa distribution enables a detailed understanding of systems far from thermal equilibrium, with applications ranging from condensed matter systems to astrophysical environments such as the interiors of neutron stars, white dwarfs and giant planets. It may also provide insights into quantum phenomena in high-energy and nano-scale systems, including electron gases, electron-hole systems in semiconductors, and nano-plasmonics and fractional quantum Hall-effect systems. This study bridges the gap between classical and quantum statistical mechanics and highlights the importance of transitioning from traditional fluid models to statistical approaches in the study of quantum plasmas, providing a significant step toward understanding mixed fermion-boson systems.

**Acknowledgment**

The authors thank SERB- DST, Govt. of India for financial support under MATRICS scheme (MTR/2021/000471).

# Figure Captions

Fig. 1  Variation of entropy ($S_\alpha^{OBF}$) with kappa index ($\kappa$), for the particle density ratio of fermions to bosons $(\Delta)$, with $T \approx 10^6 K$ and s = 5/2.

Fig. 2  Variation of distribution function ($\langle n_\alpha \rangle_{OBF}$) with energy ($\varepsilon_\alpha$), for different values of kappa index, with $T \approx 10^6 K$ and s = 5/2.

Fig. 3  Variation of partition function ($Z_{OBF}$) with temperature ($T$), for relativistic (s = 4) and non-relativistic Cases(s = 5/2), with $\kappa$ = 0.5.

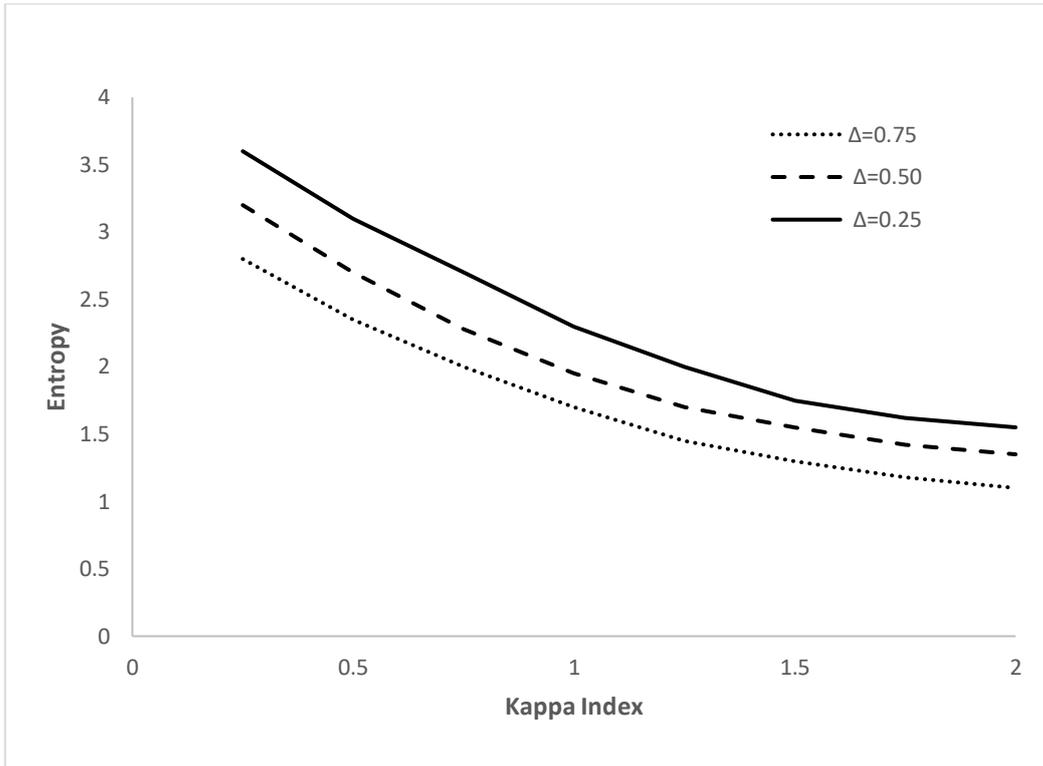

Fig. 1

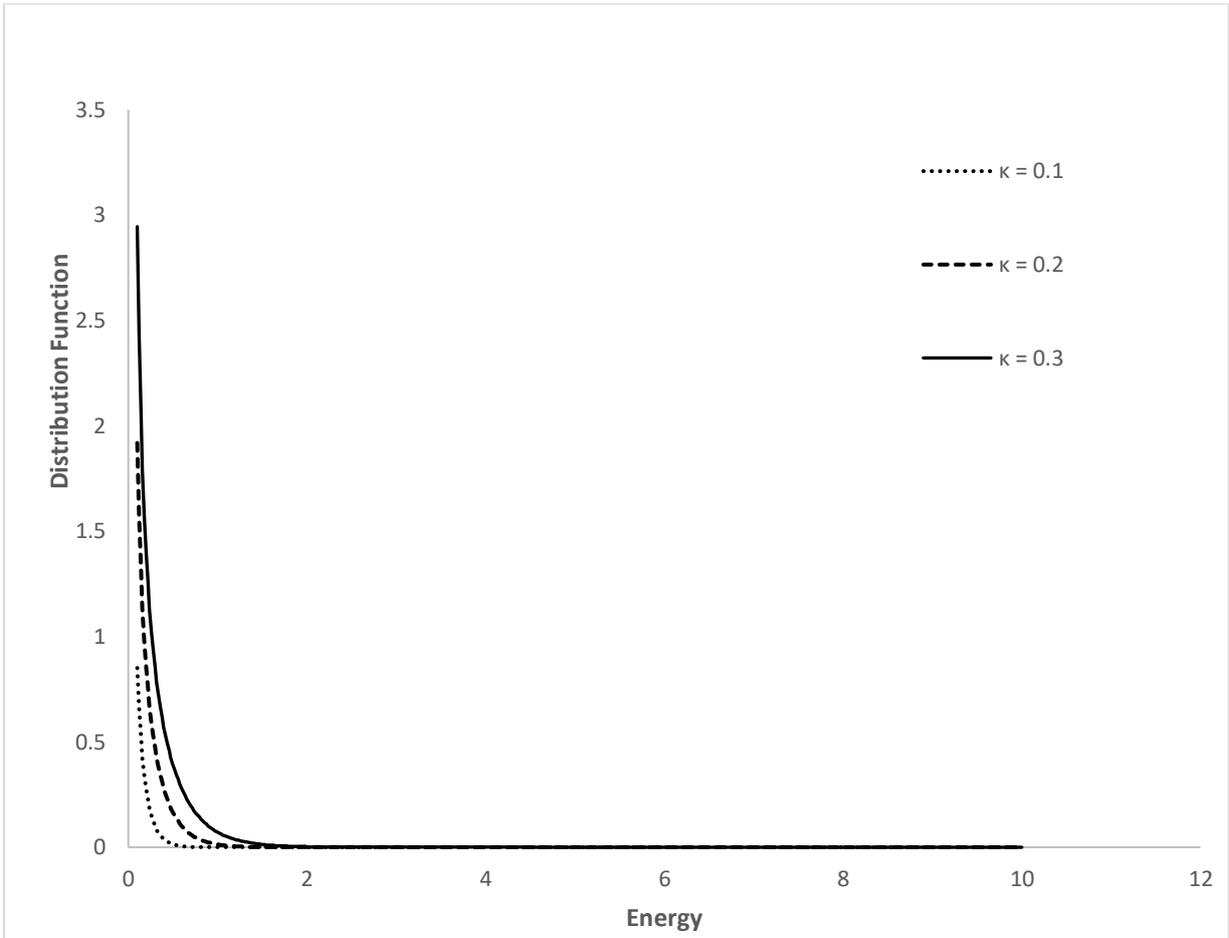

Fig. 2

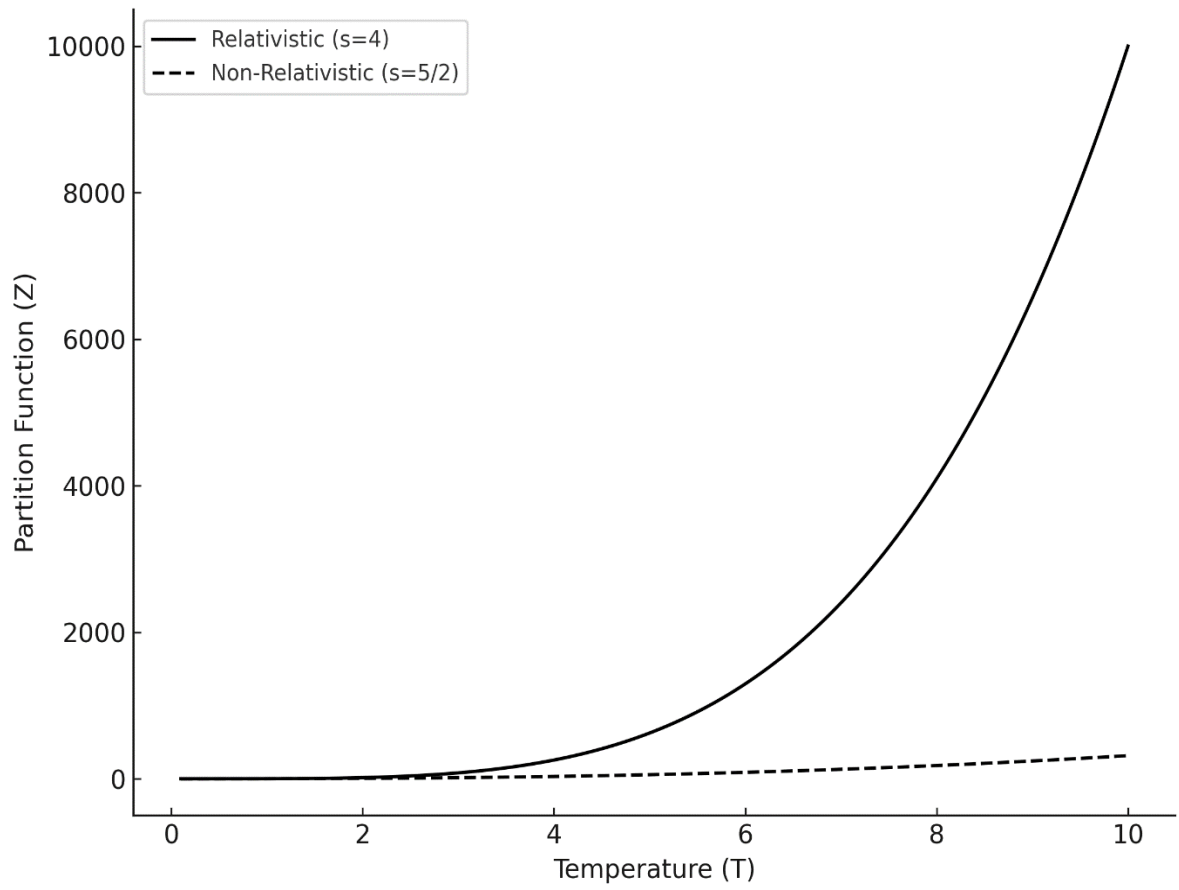

Fig. 3